\documentclass[english,aps,prb,reprint,superscriptaddress]{revtex4-1}
\usepackage[T1]{fontenc}
\usepackage[latin9]{inputenc}
\usepackage{geometry}
\usepackage{array}
\usepackage{amssymb}
\usepackage{graphicx}
\usepackage{esint}

\makeatletter

\providecommand{\tabularnewline}{\\}

\makeatother

\usepackage{babel}
\begin{document}

\title{The effects of Coulomb interactions on the superconducting gaps in
iron-based superconductors}

\author{Zhidong Leong}

\author{Philip Phillips}

\affiliation{Department of Physics and Institute for Condensed Matter Theory, University of Illinois at Urbana-Champaign,
Urbana, Illinois 61801, USA}
\begin{abstract}
Recent angle-resolved photoemission spectroscopy measurements of Co-doped LiFeAs report a large and robust superconducting gap on the $\Gamma$-centered hole band that lies 8 meV
below the Fermi level. We show that, unlike a conventional superconductor described by BCS theory, a multiband system with strong interband Coulomb interactions can explain these observations. We model LiFeAs with a five-band model in which the shallow hole band is coupled with the other bands by only Coulomb interactions. Using Eliashberg theory, we find reasonable interaction parameters that reproduce the $T_c$ and all five gaps of LiFeAs. The energy independence of the Coulomb interactions then ensures the robustness of the gap induced on the shallow band. Furthermore, due to the repulsive nature of the Coulomb interactions, the gap changes sign between the shallow band and the other hole pockets, corresponding to an unconventional $s_{\pm}$ gap symmetry. Unlike other families of iron-based superconductors, the gap symmetry of LiFeAs has not been ascertained experimentally. The experimental implications of this sign-changing state are discussed.
\end{abstract}
\maketitle

\section{Introduction}

Despite the differences between LiFeAs and other families of iron-based superconductors \cite{Tapp2008,Borisenko2010}, the superconductivity shares much in terms of family resemblance \cite{Taylor2011}. Experiments have found that the anisotropic gaps in LiFeAs predominantly arise from antiferromagnetic spin fluctuations \cite{Umezawa2012,Allan2012,Allan2014}. These spin fluctuations originate from scattering between the electron pockets located at the $M$-point and the hole pockets at the $\Gamma$-point. In addition, theoretical and first-principles calculations have confirmed the importance of spin fluctuations in LiFeAs, predicting an $s_{\pm}$ gap symmetry similar to that of many iron-based superconductors \cite{Platt2011,Wang2013e,Ummarino2013, Yin2013a}.

However, the superconductivity mechanism at the innermost hole pocket at the $\Gamma$-point of LiFeAs remains a puzzle. Since the band barely crosses the Fermi level, the small size of the pocket makes studying it challenging. Theoretical calculations have shown that spin fluctuations alone are insufficient to account for the large gap found on the tiny pocket \cite{Wang2013e}. Furthermore, unlike the other pockets, the innermost hole pocket has an isotropic gap \cite{Umezawa2012,Allan2012}, suggesting the presence of a different pairing mechanism. 

\begin{figure}
\begin{centering}
\includegraphics[width=0.95\columnwidth]{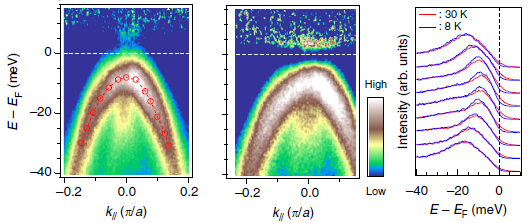}
\par\end{centering}

\caption{ARPES results of 3\% Co-doped LiFeAs from Ref. \onlinecite{Miao2015}. Left: the ARPES intensity plot in the normal state, showing that the shallow hole band at the $\Gamma$-point is 8 meV below the Fermi level. Middle: the same plot in the superconducting state, showing that the large gap on the shallow band is robust against the Lifshitz transition. Right: a combined plot of the shallow band's spectral weights in both the normal and superconducting states, showing significant changes between the two states even well below the Fermi level.
\label{fig:ARPES_results}}
\end{figure}

Recently, high-resolution angle-resolved photoemission spectroscopy (ARPES) measurements have found that the gap on the innermost hole pocket is robust, even as the shallow band sinks 8 meV below the Fermi level upon electron doping \cite{Miao2015}. As shown in Figure \ref{fig:ARPES_results}, the gap remains large, and the band's spectral weight significantly changes between the normal and superconducting states, even at energies well below the Fermi level. These observations suggest that the gap on the shallow band does not arise from low-energy excitations related to the structure of the Fermi surface.

Prior to the ARPES measurements \cite{Miao2015}, theoretical calculations have pointed to orbital-spin fluctuations \cite{Saito2014} and the renormalization by high-energy excitations \cite{Ahn2014} as means to obtain the large gap on the shallow band. However, it is unclear whether any of these proposed mechanisms will allow the large gap to be robust across the Lifshitz transition upon electron doping. 

In this paper, we propose Coulomb interactions as the superconductivity mechanism at the shallow hole band centered at the $\Gamma$-point in LiFeAs. We represent LiFeAs by a five-band model, in which the shallow band couples to the other bands by only Coulomb interactions. Using Eliashberg theory, we find that interband Coulomb interactions induce a large superconducting gap on the shallow band. The energy independence of Coulomb interactions then ensures the robustness of the gap against changes in the Fermi level, in agreement with ARPES observations \cite{Miao2015}.  Using reasonable interaction parameters, we find that our model can quantitatively reproduce the experimental values of the $T_c$ and the gaps on all five bands. Relative to the case without Coulomb interactions, these interactions are found to enhance the $T_c$ by a factor of $1.7$, indicating the significant role they play in LiFeAs. Finally, due to the repulsive nature of Coulomb interactions, our results predict an unconventional $s_{\pm}$ gap symmetry, in which the gap changes sign between the hole pockets at the $\Gamma$-point. Unlike other families of iron-based superconductors, the gap symmetry of LiFeAs has not been ascertained experimentally.

\section{Methods}

Since iron-based superconductors are moderately coupled \cite{Ding2008,Lin2008,Miao2012,He2013,Miao2015},
we employ Eliashberg theory \cite{Eliashberg1960}. Although our approach is phenomenological as in Ref. \onlinecite{Ummarino2013}, BCS theory cannot be used, because it would be unable to yield the correct gap to $T_c$ ratio.  It is for this reason that we adopt the Eliashberg approach.

In the Matsubara formalism, the multiband Eliashberg gap equations are \cite{Carbotte1990}

\begin{widetext}

\begin{eqnarray}
Z_{i}\left(i\omega_{n}\right) & = & 1+\frac{1}{2n+1}\sum_{j,m}\frac{\omega_{m}}{\sqrt{\omega_{m}^{2}+\Delta_{j}^{2}\left(i\omega_{m}\right)}}\left[V_{ij}^{\mathrm{ph}}\left(i\omega_{n}-i\omega_{m}\right)+V_{ij}^{\mathrm{sp}}\left(i\omega_{n}-i\omega_{m}\right)\right],\\
Z_{i}\left(i\omega_{n}\right)\Delta_{i}\left(i\omega_{n}\right) & = & \pi T\sum_{j,m}\frac{\Delta_{j}\left(i\omega_{m}\right)}{\sqrt{\omega_{m}^{2}+\Delta_{j}^{2}\left(i\omega_{m}\right)}}\left[V_{ij}^{\mathrm{ph}}\left(i\omega_{n}-i\omega_{m}\right)-V_{ij}^{\mathrm{sp}}\left(i\omega_{n}-i\omega_{m}\right)-\mu_{ij}\theta\left(\left|\omega_{m}\right|-\omega_{c}\right)\right].\label{eq:Eli2}
\end{eqnarray}

\end{widetext}

Here, $i,j$ are the band indices, and $\omega_{n}=\left(2n+1\right)\pi T$
is the $n$th fermionic Matsubara frequency. The function $Z_{i}$$\left(i\omega_{n}\right)$
describes corrections to the electron self-energy, and $\Delta_{i}\left(i\omega_{n}\right)$ is the energy-dependent superconducting
gap of the $i$th band. The potentials $V_{ij}^{\mathrm{ph}}$
and $V_{ij}^{\mathrm{sp}}$, defined by 
\begin{eqnarray}
V_{ij}^{\mathrm{ph,sp}}\left(iq_{m}\right) & = & 2\int_{0}^{\infty}\omega d\omega\frac{\alpha^{2}F_{ij}^{\mathrm{ph,sp}}\left(\omega\right)}{\omega^{2}+q_{m}^{2}},
\end{eqnarray}
describes the effects of band $i$ on band $j$ due to interactions
with phonons and spin fluctuations, respectively. The Eliashberg functions
$\alpha^{2}F_{ij}\left(\omega\right)$ can be experimentally determined
from the inversion of tunneling data. In Eq. \ref{eq:Eli2}, notice
that the potential mediated by spin fluctuations appears with a negative
sign. This is due to the spin-flip nature of magnon scattering. Finally,
the pseudopotential $\mu_{ij}$ describes the Coulomb interactions
between bands $i$ and $j$. It is usually given a cutoff at a
large energy $\omega_{c}$ for numerical convergence. Physically,
the cutoff signifies the existence of an energy scale up to which the Coulomb interaction is
instantaneous. Unlike boson-mediated
interactions, the Coulomb interactions cannot be easily measured,
and are usually tuned phenomenologically to fit experimental measurements.

For a given set of interaction parameters, we self-consistently
solve the Eliashberg equations for the superconducting gaps $\Delta_{i}\left(i\omega_{n}\right)$.
Then, using Pad\'e approximants \cite{Vidberg1977}, we perform an analytic
continuation to obtain the solutions in terms of real energies. The
energy gap $\Delta_{0}$ measured in ARPES experiments is then given
by $\Delta_{0}=\mbox{Re}\left[\Delta\left(\omega=\Delta_{0}\right)\right]$.

\section{Five-band model}

\begin{figure}
\begin{centering}
\includegraphics[width=0.95\columnwidth]{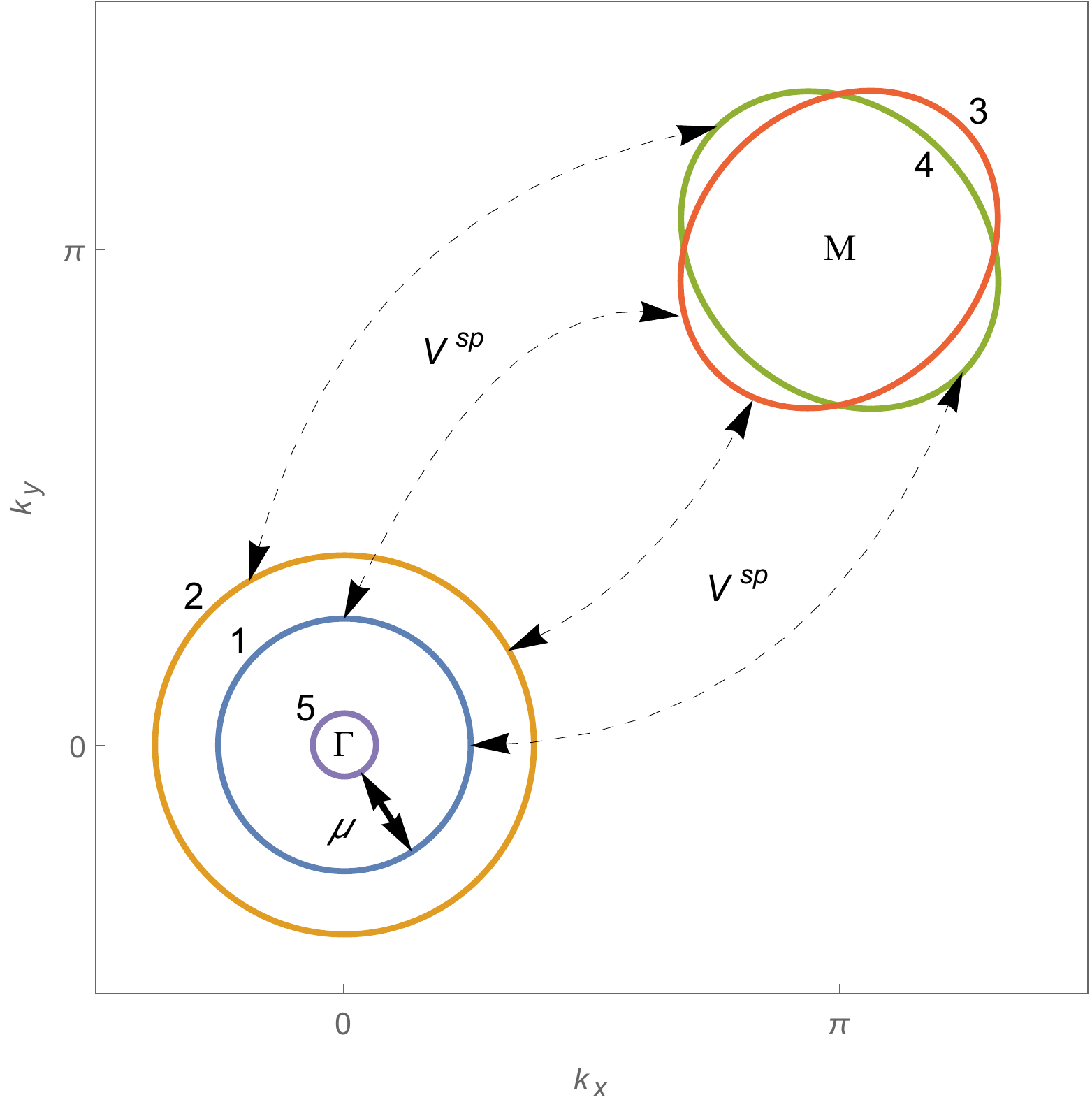}
\par\end{centering}

\caption{A schematic of the Fermi surfaces of LiFeAs. In our model, the deep
hole bands $\left(1,2\right)$ are coupled with the deep electron
bands $\left(3,4\right)$ via interactions mediated by spin-fluctuations.
In addition, the shallow hole band $\left(5\right)$ is coupled with
the inner deep hole band $\left(1\right)$ via Coulomb interactions.
\label{fig:fermi_surface}}
\end{figure}

We represent LiFeAs by a five-band model with a Fermi surface schematically shown in Figure \ref{fig:fermi_surface}. The deep bands $(1, 2)$ at the $\Gamma$-point 
are coupled to the deep bands $(3,4)$ at the $M$-point by interband interactions mediated by spin-fluctuations,
while the shallow band $(5)$ at the $\Gamma$-point is coupled to the deep bands $(1, 2, 3, 4)$ by Coulomb interactions. As in previous studies \cite{Wang2013e,Ummarino2013,Saito2014,Ahn2014}, the spin fluctuations arise from interband scattering between the hole and electron pockets.
Phonon-mediated interactions involving only the deep bands are assumed
to be negated by Coulomb interactions acting likewise, and boson-mediated
interactions involving the shallow band are assumed to be negligible
due to the low density of states at the Fermi level.

Previously, a four-band model of LiFeAs was studied under similar
assumptions \cite{Ummarino2013}. The authors omitted the shallow
hole band  claiming that its small density of states at the Fermi
level precludes its contribution to superconductivity. They found
that a large intraband phonon-mediated interaction on band 1 at the $\Gamma$-point is required to quantitatively reproduce the superconducting gaps
of LiFeAs. Here, we will show that this interaction is not necessary
in our full five-band model.

For the spin fluctuations in our model, we follow Ref. \onlinecite{Ummarino2013}
and use Lorentzian Eliashberg functions $\alpha^{2}F_{ij}^{\mathrm{sp}}\left(\omega\right)$
with peak energies $\Omega_{ij}=8\mbox{\ meV}$ and half-widths $Y_{ij}=4\mbox{\ meV}$.
The coupling constants 
\begin{eqnarray}
\lambda_{ij} & = & \left(\begin{array}{ccccc}
0 & 0 & \lambda_{13} & \lambda_{14} & 0\\
0 & 0 & \lambda_{23} & \lambda_{24} & 0\\
\lambda_{31} & \lambda_{32} & 0 & 0 & 0\\
\lambda_{41} & \lambda_{42} & 0 & 0 & 0\\
0 & 0 & 0 & 0 & 0
\end{array}\right),
\end{eqnarray}
defined by 
\begin{eqnarray}
\lambda_{ij} & = & 2\int_{0}^{\infty}d\Omega\frac{\alpha^{2}F_{ij}^{\mathrm{sp}}\left(\Omega\right)}{\Omega},
\end{eqnarray}
satisfy $\lambda_{31}/\lambda_{13}=0.9019$, $\lambda_{41}/\lambda_{14}=1.5010$,
$\lambda_{32}/\lambda_{23}=1.0483$, and $\lambda_{42}/\lambda_{24}=1.7447$,
in accordance with band structure calculations \cite{Ummarino2013}. 

For the Coulomb interactions, since they are generally stronger for smaller momentum transfers, we include them only between the two innermost hole bands, that is only $\mu_{15},\mu_{51}\neq0$. This can also be justified by the fact that only the two inner hole bands have similar orbital content \cite{Ahn2014}. Furthermore, intraband Coulomb interactions $\mu_{55}$ can be omitted, as has been shown in a functional renormalization analysis \cite{Platt2011}. While the values of $\mu_{ij}$ can be calculated from first principles, doing so is
difficult, as they depend on the details of the band structure. The cutoff energy $\omega_{c}$ of the Coulomb interactions is set to be $10\Omega_{ij}$, as is commonly done in the literature \cite{Carbotte1990}.

\section{Results}

\begin{table}
\begin{tabular}{|>{\centering}p{0.15\columnwidth}|>{\centering}p{0.15\columnwidth}||>{\centering}p{0.15\columnwidth}|>{\centering}p{0.15\columnwidth}||>{\centering}p{0.15\columnwidth}|c|}
\hline 
\multicolumn{2}{|>{\centering}m{0.3\columnwidth}||}{Interaction parameters} & \multicolumn{2}{>{\centering}m{0.3\columnwidth}||}{Energy gaps / meV} & \multicolumn{2}{>{\centering}m{0.3\columnwidth}|}{$T_{c}$ / K}\tabularnewline
\hline 
\hline 
$\lambda_{13}$ & 1.05 & $\Delta_{1}$ & 5.0 & Expt. & 18.0\tabularnewline
\hline 
$\lambda_{23}$ & 0.73 & $\Delta_{2}$ & 2.6 & Model & 20.4\tabularnewline
\hline 
$\lambda_{14}$ & 0.41 & $\Delta_{3}$ & -3.6 &  & \tabularnewline
\hline 
$\lambda_{24}$ & 0.30 & $\Delta_{4}$ & -2.9 &  & \tabularnewline
\hline 
$\mu_{51}$ & 0.38 & $\Delta_{5}$ & -5.5 &  & \tabularnewline
\hline 
\end{tabular}

\caption{The unique values of the interaction parameters $\lambda_{13},\lambda_{14},\lambda_{23},\lambda_{24},\mu_{15}$
used to reproduce the five superconducting gaps at low temperatures.
The $T_{c}$ resulting from this set of parameters is consistent with
experimental measurements. \label{tab:param}}
\end{table}

Although our model has six adjustable parameters $\lambda_{13},\lambda_{14},\lambda_{23},\lambda_{24},\mu_{15},$
and $\mu_{51}$, we find that reproducing the five gaps of LiFeAs
requires $\mu_{15}\gtrsim0.2$. We adopt the minimum value here to obtain a lower bound on the effects of Coulomb interactions. Now that the model is left
with five adjustable parameters, they can be uniquely solved to reproduce
the five superconducting gaps at low temperatures. The results are
shown in Table \ref{tab:param}. 

\begin{figure}
\begin{centering}
\includegraphics[width=0.95\columnwidth]{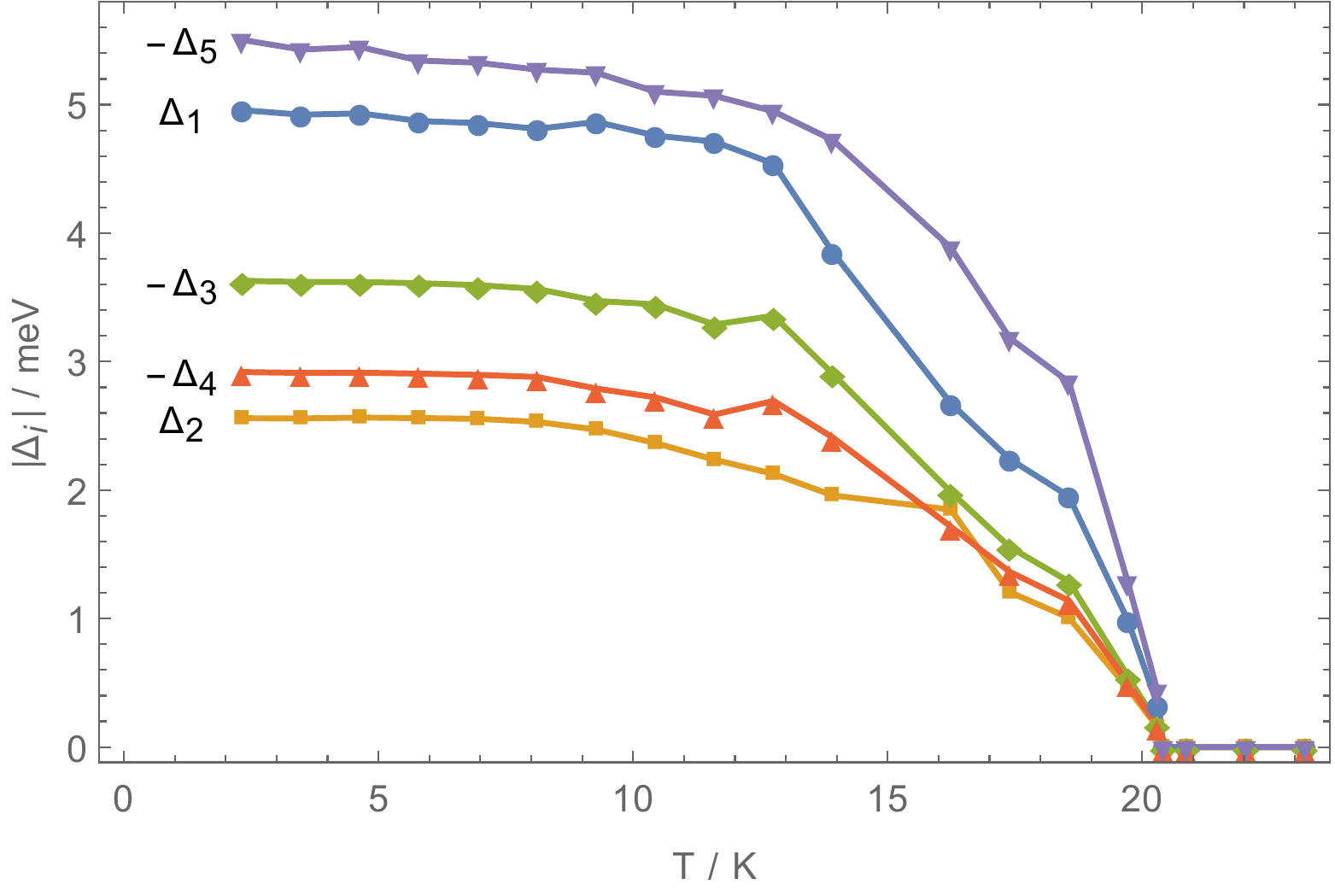}
\par\end{centering}

\caption{A plot of the energy gap against temperature for the five-band model
of LiFeAs. The gaps agree with experimental measurements at low temperatures.
The interaction parameters used are shown in Table \ref{tab:param}.
\label{fig:gap_vs_T}}
\end{figure}

Using these parameters, we calculated the gaps at various temperatures, as shown in Figure \ref{fig:gap_vs_T}. The temperature dependence has the expected mean-field form with a $T_{c}\approx20.4$ that is consistent with experimental measurements. Such reproduction of the $T_c$ in addition to the gaps is often omitted in theoretical calculations. 
The larger errors at higher temperatures are expected, due to the reduced number of Matsubara points. Notice that the opposite signs between $\Delta_{1},\Delta_{2}$
and $\Delta_{3},\Delta_{4}$ are due to the spin-flip nature of the interactions mediated by spin-fluctuations, while the opposite signs between $\Delta_{1}$ and $\Delta_{5}$ are due to the repulsive nature of Coulomb interactions.

\begin{figure}
\begin{centering}
\includegraphics[width=0.95\columnwidth]{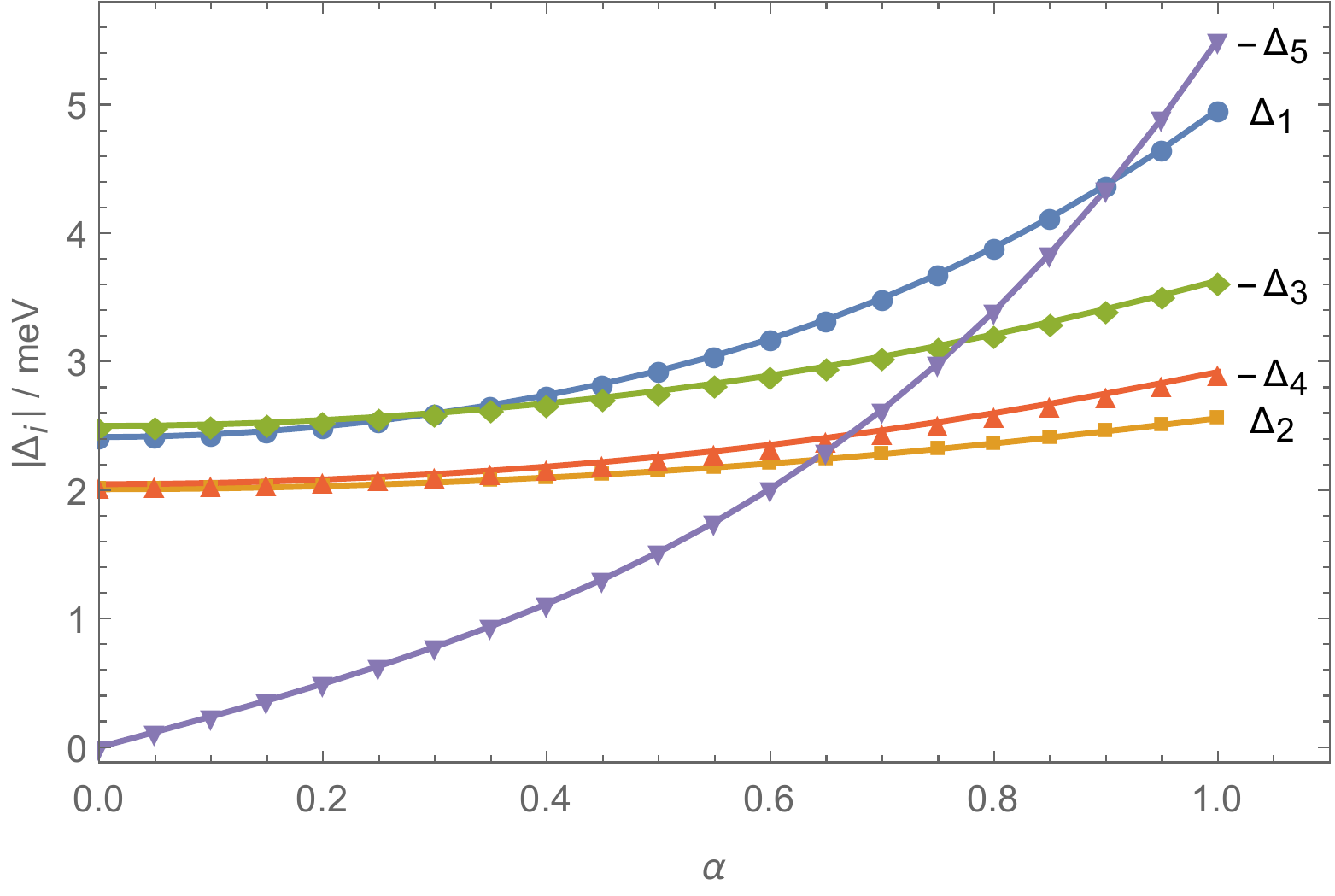}
\par\end{centering}

\caption{A plot of the energy gaps against the Coulomb interactions showing
Coulomb interactions inducing a large gap on the shallow hole band.
Due to the energy independence of Coulomb interactions, the induced
gap is robust against changes in the Fermi energy. \label{fig:gap_vs_mu}}
\end{figure}

To elucidate the effects of Coulomb interactions, we also performed
low temperature calculations considering cases in which the Coulomb interactions
$\mu_{ij}$ are scaled by a factor of $0\leq\alpha\leq1$. Figure \ref{fig:gap_vs_mu}
shows that in the absence of Coulomb interactions at $\alpha=0$,
the shallow hole band at the $\Gamma$-point is not gapped, $\Delta_{5}=0$. As $\alpha$
increases, interband Coulomb interactions induce a gap on the shallow
hole band, and increase the magnitude of the gaps on the other bands.
For large enough $\alpha$'s, the gap on the shallow band is the largest
within the whole Brillouin zone. Then, because Coulomb interactions
are energy independent in the regime of interest, varying only on
the energy scale of the plasma frequency $\omega_{p}\sim1\ \mbox{eV}$
\cite{Drechsler2010}, the large gap on the shallow hole band is robust
even as the band sinks below the Fermi level, in agreement
with ARPES observations \cite{Miao2015}.

\begin{figure}
\begin{centering}
\includegraphics[width=0.95\columnwidth]{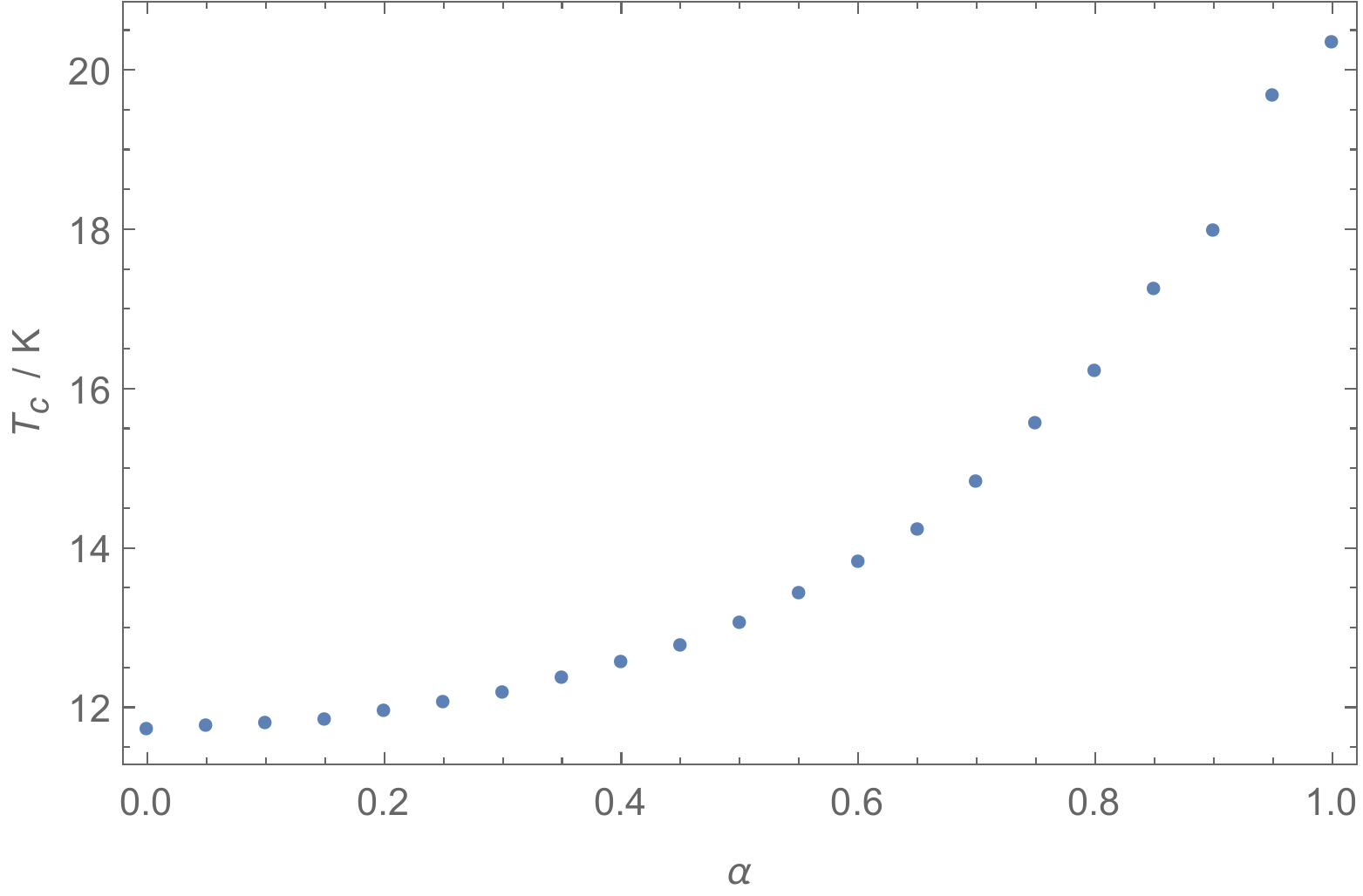}
\par\end{centering}

\caption{A plot showing the effects of Coulomb interactions on $T_{c}$. In
LiFeAs, Coulomb interactions enhance the $T_{c}$ by a factor of $1.7$.
\label{fig:Tc_enhancement_vs_mu}}
\end{figure}

Next, Figure \ref{fig:Tc_enhancement_vs_mu} shows the effects of Coulomb interactions on $T_{c}$. Relative to the case without Coulomb interactions, these interactions are found to enhance the $T_c$ by a factor of $1.7$, indicating the significant role they play in LiFeAs.


\section{Discussion and conclusions}

While Coulomb interactions are often thought to oppose superconductivity,
our results illustrate the importance of interband Coulomb interactions
as a mechanism of $T_{c}$ enhancement. Such enhancement is likely
to occur in systems with bands that have small densities of states
at the Fermi level as Coulomb interactions dominate pairing for these
bands. An example of such systems is single-layer FeSe, which has
a hole band completely below the Fermi level \cite{Liu2012c}. Pairing by interband
Coulomb interactions may be important in the search of new high-$T_{c}$
superconductors.

Our results also predict that the superconducting gap changes sign between the inner hole pockets at the $\Gamma$-point. The resulting sign-reversal gap symmetry is unlike the conventional $s_{\pm}$ gap symmetry \cite{Mazin2008} believed to be present in iron-based superconductors. Similar unconventional gap symmetries have also been proposed in Ref. \onlinecite{Ahn2014}. Unlike other iron-based superconductors in which the pairing symmetry has been extensively studied \cite{Paglione2010}, such is not the case for LiFeAs. Measurement of the gap symmetry can be performed using the SQUID junction proposed in Ref. \onlinecite{Wu2009}. While there have been experiments \cite{Chi2014} measuring the gap symmetry of LiFeAs, the results are not conclusive \cite{Ahn2014}, as the different hole bands were not individually resolved.

ARPES experiments \cite{Lubashevsky2012,Okazaki2014} have also suggested that some iron-based superconductors are in the BCS-BEC (Bose-Einstein condensate) crossover regime, due to the large ratios of the superconducting gap to the chemical potential observed. This observation can alternatively be understood in the context of our results. Since Coulomb interactions are energy independent, they can yield the observed superconducting gaps regardless of the size of the chemical potential, thereby providing a mechanism for the gapping of states lying below the Fermi energy.

Recent experiments have provided further evidence that superconductivity in LiFeAs does not entirely arise from low-energy spin fluctuations. Nuclear magnetic resonance (NMR) measurements show that the $T_c$ in Co-doped LiFeAs decreases even when the strength of spin-fluctuations increases upon doping \cite{Dai2015b}. Furthermore, tunneling spectroscopy measurements show the existence of a temperature independent bosonic mode not directly related to spin fluctuations \cite{Nag2015}. These results strengthen our case that Coulomb interactions are important for superconductivity in LiFeAs.

Before we conclude, we would like to highlight a common misconception about multiband superconductivity found in the discussions of the ARPES results in Ref. \onlinecite{Miao2015}.  Unlike a one-band system, a gap $\Delta_i$ driven by interband interactions in a multiband system depends on the density of states $N_j$ of the other bands, and not on its own density of state $N_i$. For example, in a two-band system with only interband interactions, the ratio $\Delta_1 / \Delta_2$ of the gaps is proportional to $\sqrt{N_2/N_1}$ \cite{Dolgov2009}. Consequently, the shallow band in LiFeAs developing a large superconducting gap does not contradict the principles of BCS theory, as was incorrectly implied in Ref. \onlinecite{Miao2015}. Nevertheless, the paper's main arguments remain sound.

In conclusion, we proposed Coulomb interactions as the superconductivity mechanism at the shallow hole band centered at the $\Gamma$-point in LiFeAs. We represented LiFeAs by a five-band model, in which the shallow band couples to the other bands by only Coulomb interactions. Using Eliashberg theory, we found that interband Coulomb interactions can induce a large superconducting gap on the hole band. The energy independence of Coulomb interactions then ensures the robustness of the gap, in agreement with the ARPES observations \cite{Miao2015}. Using reasonable interaction parameters, we found that our model can quantitatively reproduce the experimental values of  $T_c$ and the gaps on all five bands. Relative to the case without Coulomb interactions, these interactions were found to enhance the $T_c$ by a factor of $1.7$, indicating the significant role they play in LiFeAs. Finally, due to the repulsive nature of Coulomb interactions, our results predict an unconventional $s_{\pm}$ gap symmetry, in which the gap changes sign between the hole pockets at the $\Gamma$-point. This study should help motivate further experiments on the pairing symmetry of LiFeAs.

\begin{acknowledgments}
Z. Leong is supported by a scholarship from the Agency of Science,
Technology and Research. P. Phillips is supported by the Center for
Emergent Superconductivity, a DOE Energy Frontier Research Center,
Grant No. DE-AC0298CH1088 and the J. S. Guggenheim Foundation.
\end{acknowledgments}
\bibliographystyle{apsrev4-1}
\bibliography{library}

\end{document}